\documentclass[11pt]{article}

\usepackage[utf8]{inputenc}
\usepackage[T1]{fontenc}
\usepackage{amsmath,amssymb}
\usepackage{graphicx}
\usepackage{booktabs}
\usepackage{hyperref}
\usepackage[margin=1in]{geometry}
\usepackage{natbib}
\usepackage{lineno}
\usepackage{xcolor}
\usepackage{float}
\usepackage{multirow}
\usepackage{array}
\usepackage{caption}
\usepackage{subcaption}
\usepackage{setspace}
\usepackage{threeparttable}

\title{\textbf{MCAnalysis: An Open-Source Package for Preprocessing, Modelling,
and Visualisation of Menstrual Cycle Effects in Digital Health Data}}

\author{
  Kyra Delray\textsuperscript{1,*},
  Glyn Lewis\textsuperscript{2},
  Bola Grace\textsuperscript{3},
  Joseph Hayes\textsuperscript{2} \&
  Robin Evans\textsuperscript{1} \\[6pt]
  \textsuperscript{1}Department of Statistics, University of Oxford \\
  \textsuperscript{2}Division of Psychiatry, University College London \\
  \textsuperscript{3}Institute for Women's Health, University College London \\
  \textsuperscript{*}Corresponding author
}

\date{}

\begin{document}

\maketitle

\begin{abstract}

\textbf{Background:} The menstrual cycle influences numerous physiological and
psychological outcomes, yet standardised, open-source statistical methods for
quantifying these cyclic effects remain lacking. Existing approaches often fail
to account for the inherent periodicity of cycle data, lack integrated
preprocessing pipelines, or require specialist statistical expertise, limiting
reproducibility across studies.

\textbf{Methods:} We developed \texttt{mcanalysis}, an open-source statistical
package available in both R and Python, implementing a Fourier-basis generalised
additive model (GAM) specifically designed for menstrual cycle research. The
package provides a complete analytical pipeline: processing period dates,
labelling cycle days relative to menstruation onset (Day~0), filtering
physiologically plausible cycles (21--35 days), normalising outcomes to
individual means, fitting cyclic GAMs with bootstrap confidence intervals, and
identifying turning points to generate phase-specific linear trend estimates. We
demonstrate the utility of the package on 15 wearable and self-reported outcomes
across physiological, sleep, symptom, mood, and activity domains, using data
from the Juli chronic health management application ($N = 2{,}816$ users with
recorded menstruation dates).

\textbf{Results:} Nine of 15 outcomes showed evidence of association with the
menstrual cycle ($p < 0.05$). Heart rate variability showed the strongest
evidence of a cycle effect ($p < 0.001$), with a
trough at Day~$-7$ (pre-menstrual) and peak at Day~$+4$ (post-menstrual).
Oxygen saturation also showed evidence of a cycle effect ($p = 0.002$), with a
novel multi-peak pattern. Sleep duration showed evidence of a cycle effect
($p = 0.003$). Migraine ($p < 0.001$) and headache ($p =
0.005$) showed cycle effects consistent with known patterns of menstrual
migraine. Among mental health outcomes, EMA mood ($p = 0.024$), PHQ-8 lack of
energy ($p = 0.008$), and mania scores ($p =
0.041$) showed evidence of cycle effects, though
the latter two are based on small samples and require replication. Hours outside
also showed evidence of a cycle effect ($p = 0.019$). No tested confounders
(age, mean steps, mean sleep) were significantly associated with
cycle-normalised outcomes.

\textbf{Conclusions:} \texttt{mcanalysis} provides a standardised, reproducible,
comprehensive approach to menstrual cycle analysis for users at all levels of
statistical expertise. Detectable cycle effects across physiological, sleep,
symptom, mood, and activity domains suggest that cycle-aware adjustment of
health recommendations and digital health interventions may benefit users across
a broad range of applications. The cyclic structure of the fitted models makes
outputs directly integrable into personalised intervention frameworks such as
dynamic treatment regimes. The package is freely available at
\url{https://github.com/kyradelray/mcanalysis}, with a no-code web interface at
\url{https://kyradelray.shinyapps.io/mcanalysis/}.

\end{abstract}

\clearpage

\section{Introduction}

Digital Health Technologies (DHTs) including consumer wearable devices and
digital health applications offer an opportunity for continuous, large-scale
data collection. Wearables give insight into physiological biomarkers that help
us understand the human body, through passive data collection. Such data can be
collected at a regularity that would be impossible otherwise. Digital health
applications provide the chance to collect diverse types of data from clinically
validated surveys, GPS, and contextual inputs. This combination has the ability
to make profound advances in our understanding of the factors that affect
individuals on a personal and population level~\citep{grace2025digital}.

One of these factors is the menstrual cycle. Particularly because of its
inter-individual variability, studying it requires large sample sizes, and to
truly grasp its effects on the human body, it needs to be observed on a
near-daily scale~\citep{bull2019real}. The menstrual cycle, typically lasting
21--35 days, is governed by fluctuations in oestrogen and progesterone that
influence cardiovascular, autonomic, metabolic, and psychological
parameters~\citep{baker2007circadian}. Studies have identified
cycle-phase-dependent changes in resting heart rate (RHR), heart rate
variability (HRV), body temperature, sleep, and
mood~\citep{delray2025mood, jasinski2024cardiovascular, shilaih2018modern,
maijala2019nocturnal, grant2020menstrual, goodale2019wearable,
dezambotti2015menstrual, soumpasis2020real}.

These fluctuations represent a source of predictable, learnable variation:
quantifying them improves model accuracy in population health research, supports
personalised clinical monitoring, and empowers individuals to understand
otherwise unexplained day-to-day changes in their own physiology and mood.

Despite growing interest in understanding the effects of the menstrual cycle on
mental and physiological factors, the field lacks standardised statistical
methods for menstrual cycle analysis. Common approaches, such as comparing means
across coarsely defined phases or fitting standard regression models, fail to
account for the inherent periodicity of cycle data, where Day~$-14$ and
Day~$+14$ represent the same biological timepoint (the day before menstruation,
whether approached from the prior or current
cycle)~\citep{schmalenberger2021menstrual}. Furthermore, the analytical pipeline
from raw data to publication-ready results requires multiple preprocessing steps
(cycle identification, quality filtering, within-person normalisation) that are
rarely standardised across studies, hindering reproducibility and cross-study
comparison~\citep{symul2019assessment}. Finally, existing methods typically
produce outputs, such as unconnected and non-cyclical fitted
curves~\citep{jasinski2024cardiovascular} or phase
means~\citep{sims2021patterns}, that may not be intuitively interpretable for
individuals seeking to understand their own data and are difficult for clinicians
making treatment decisions, and therefore lack utility in real-world
applications.

Several statistical approaches have been used in menstrual cycle models
including phase-based comparison methods such as t-tests and ANOVA as well as
semi-parametric methods such as spline-based methods and GAMs alongside
multilevel models~\citep{cornelissen2014cosinor, wood2017generalized}. However,
these methods are typically implemented as one-off analysis scripts rather than
reusable software, lack integrated preprocessing, and are seldom validated
across different types of outcomes. To our knowledge, no open-source package
exists that provides a complete, end-to-end pipeline, from raw period dates and
outcome data through preprocessing, statistical modelling, turning point
detection, phase analysis, and publication-ready visualisation, in both R and
Python.

Here we present \texttt{mcanalysis}, an open-source package that addresses this
gap. The package implements a Fourier-basis generalised additive model
(GAM)~\citep{wood2017generalized} with two harmonics, ensuring cyclic continuity
while capturing non-sinusoidal patterns in the data. We demonstrate the package
on the Juli dataset ($N = 2{,}816$ users with recorded period dates and a range
of chronic health conditions; 15 wearable and self-reported outcomes),
identifying 9 outcomes with evidence of cycle effects across physiological,
sleep, symptom, mood, and activity domains.

\section{Methods}

\subsection{Data source: Juli dataset}

The Juli dataset comprised wearable and self-reported data for outcomes as well
as menstruation dates from 2,816 users of the Juli chronic health management
tracking application, with self-reported diagnoses of various conditions such as
depression, asthma, or bipolar disorder~\citep{grace2025digital}. Fifteen
outcome variables spanned physiological measures (HRV, oxygen saturation),
sleep (time asleep), logged symptoms (headache, pain, migraine, asthma),
lifestyle factors (hours outside), activity measures (steps, active energy,
basal energy), and mental health outcomes (EMA mood, EMA energy, PHQ-8 lack of
energy, mania).

Users also logged daily mood and energy ratings via a daily ecological momentary
assessment (EMA), rating their perceived mood and energy levels on a 2$\times$2
grid summarised to a 0--10 scale for analysis. These were found to be
statistically related to the clinical depression questionnaire PHQ-8 in previous
work~\citep{delray2025mood}. Biweekly PHQ-8 scores and representations of mania
were also collected. User-level confounder data included age, mean daily steps,
and mean sleep duration.

\subsection{The mcanalysis package}

\texttt{mcanalysis} is an open-source statistical package implemented in both R
and Python, available at \url{https://github.com/kyradelray/mcanalysis}. A
web-based interface requiring no programming is also available at
\url{https://kyradelray.shinyapps.io/mcanalysis/}. Both R and Python
implementations produce identical numerical results and follow the same
analytical pipeline. The package is designed as a single-entry-point tool:
users supply period dates, outcome data, and optional confounder data, and the
package returns statistical results and publication-ready figures.

The R and Python packages provide programmatic access to the full analytical
pipeline, including:
\begin{itemize}
  \item Preprocessing functions for cycle identification, filtering, and
    normalisation
  \item GAM fitting with bootstrap confidence intervals
  \item Turning point detection and phase-specific linear models
  \item Confounder analysis: estimation of confounder effects and fitting of
    adjusted models controlling for multiple covariates
  \item Effect modifier analysis: stratified GAM fitting to examine whether
    cycle effects vary across subgroups defined by user characteristics
  \item Publication-ready visualisation with customisable aesthetics
\end{itemize}

\begin{figure}[H]
\centering
\includegraphics[width=0.95\textwidth]{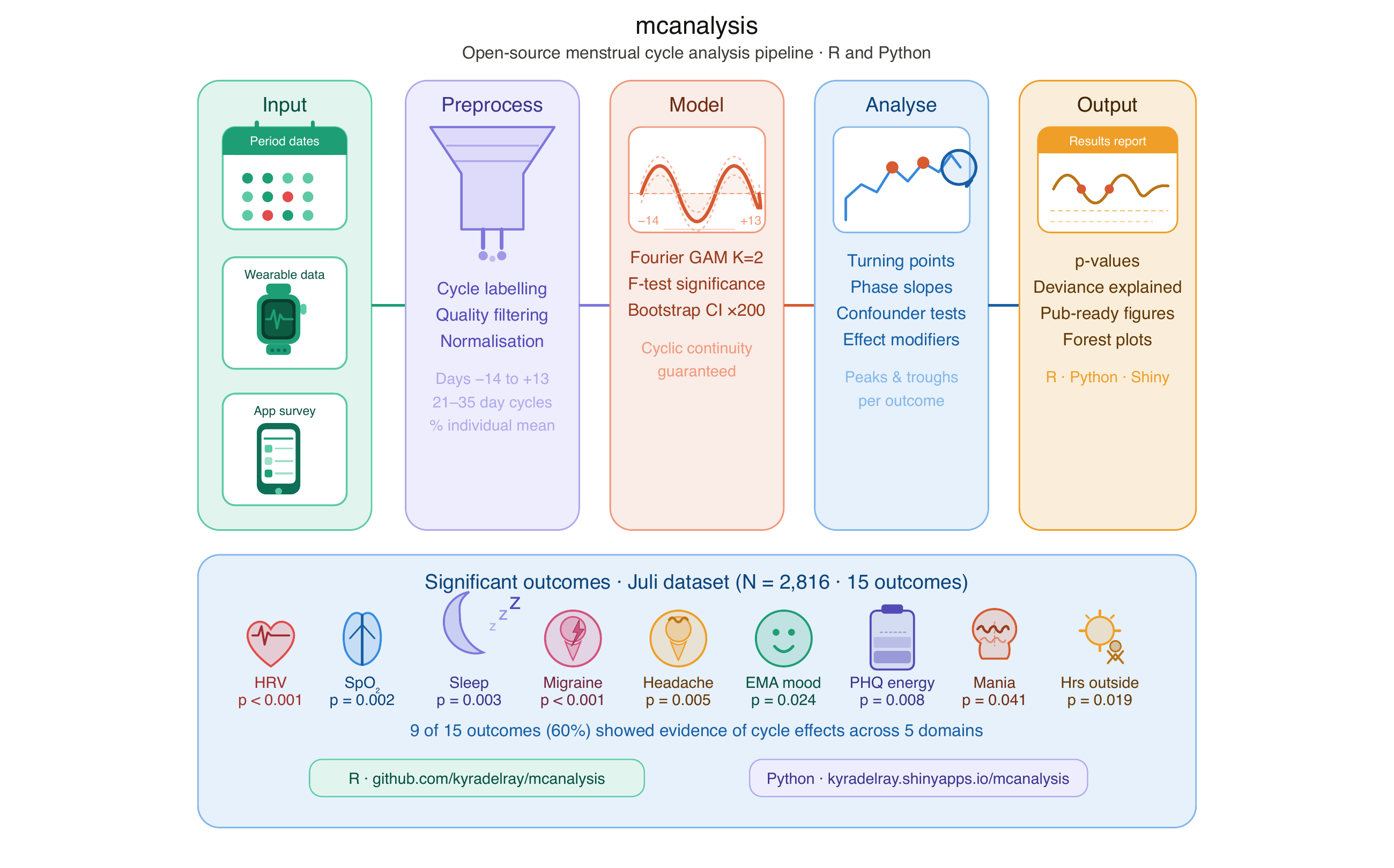}
\caption{Overview of the \texttt{mcanalysis} analytical pipeline. The package
accepts period dates, outcome data, and optional confounder data as inputs and
produces statistical results, turning point estimates, phase-specific slopes,
and publication-ready visualisations as outputs. All steps are implemented
identically in both R and Python.}
\label{fig:pipeline}
\end{figure}

The following subsections describe each stage of the pipeline
(Figure~\ref{fig:pipeline}).

\subsection{Web-based interface}
\label{sec:webapp}

For researchers without programming expertise, a web-based interface is
available at \url{https://kyradelray.shinyapps.io/mcanalysis/}. The interface
provides full access to the \texttt{mcanalysis} analytical pipeline through a
point-and-click interface, including:

\begin{itemize}
  \item \textbf{Data upload:} Users upload CSV files containing period dates
    and outcome data. The interface automatically detects column names and
    validates data format.
  \item \textbf{Interactive visualisation:} Results are displayed as
    interactive plots showing the fitted GAM curve, confidence intervals,
    turning points, and phase-specific linear models. Users can customise
    display options including raw data overlay and phase shading.
  \item \textbf{Exploratory data analysis:} Summary statistics, cycle length
    distributions, and data quality metrics are computed automatically to help
    users assess their data before modelling.
  \item \textbf{Confounder adjustment:} Users can upload confounder data and
    select variables to include in adjusted models. The interface displays both
    unadjusted and adjusted cycle curves for comparison.
  \item \textbf{Effect modifier analysis:} Users can stratify analyses by
    categorical or continuous variables to examine whether cycle effects differ
    across subgroups. For continuous modifiers, groupings are computed by
    $<$25\% quantile, middle 50\%, and top 25\%, and displayed automatically.
  \item \textbf{Synthetic example data:} A built-in example dataset with
    realistic cycle patterns allows users to explore the interface and test the
    analytical pipeline without uploading their own data.
\end{itemize}

The web interface is intended for exploratory analyses and smaller datasets.
For large-scale analyses or integration into automated pipelines, the R and
Python packages are recommended.

\subsection{Preprocessing}
\label{sec:preprocessing}

\subsubsection{Period date processing and cycle day labelling}
Period start dates are used to define menstrual cycles. Each observation is
assigned a \emph{cycle day} relative to the nearest menstruation onset, with
Day~0 defined as the first day of menstrual
bleeding~\citep{bull2019real, fehring2006variability}. Cycle days are encoded
on a centred scale from $-14$ to $+13$, such that negative days precede
menstruation (luteal phase of the prior cycle) and positive days follow
menstruation (follicular phase trending toward next menstruation).

\subsubsection{Cycle length filtering}
Cycles shorter than 21 days or longer than 35 days are excluded as
physiologically implausible or indicative of anovulatory cycles, consistent
with clinical guidelines~\citep{munro2012figo, bull2019real}.

\subsubsection{User quality filtering}
Users are retained only if they contribute at least 5 observations in each
assumed phase (negative days and positive days), ensuring sufficient data for
within-person pattern estimation.

\subsection{Normalisation}
\label{sec:normalization}

Between-person variability in outcome levels (e.g., baseline HRV differences
across individuals) can obscure within-person cycle effects. The package
normalises each user's outcome values to a percentage of their individual mean:
\begin{equation}
  y_{i,\text{norm}} = \frac{y_i}{\bar{y}_{\text{user}}} \times 100
\end{equation}
where $y_i$ is the raw outcome value and $\bar{y}_{\text{user}}$ is the user's
mean outcome across all included observations. After normalisation, 100\%
represents each user's average, and deviations reflect within-person cycle
effects in relative terms.

\subsection{Statistical modelling}
\label{sec:modelling}

\subsubsection{Fourier-basis GAM}
The package fits a cyclic generalised additive model using a Fourier basis
representation with $K = 2$ harmonics~\citep{cornelissen2014cosinor,
wood2017generalized}:
\begin{equation}
  y_i = \beta_0 + \sum_{k=1}^{K} \left[ \beta_{2k-1}
  \sin\!\left(\frac{2\pi k \cdot d_i}{28}\right) + \beta_{2k}
  \cos\!\left(\frac{2\pi k \cdot d_i}{28}\right) \right] + \varepsilon_i
  \label{eq:gam}
\end{equation}
where $y_i$ is the normalised outcome (percentage of individual mean), $d_i$ is
the cycle day ($-14$ to $+13$), $K = 2$ is the number of harmonics, and
$\varepsilon_i$ is the residual error. The Fourier basis guarantees cyclic
continuity (the curve at Day~$-14$ equals the curve at Day~$+14$) while
providing sufficient flexibility to capture non-sinusoidal waveforms. The use
of two harmonics (four basis functions plus intercept) balances model
flexibility against overfitting, allowing the model to capture asymmetric
rise--fall patterns. The Fourier basis functions are mutually orthogonal,
satisfying the no-multicollinearity assumption by construction. The within-user
bootstrap resampling scheme for confidence intervals accounts for the
non-independence of repeated observations from the same individual.

Although individual cycle lengths range from 21--35 days in the analysed
sample, the Fourier basis is fixed at a 28-day period, corresponding to the
population modal cycle length. Simulation studies have shown that
cosinor-based estimates are robust to moderate deviations in period length when
cycles are centred on menstruation onset~\citep{cornelissen2014cosinor}, and the
within-person normalisation ensures that individual baseline differences do not
confound the shape estimate.

\subsubsection{Significance testing}
\label{sec:significance}
An $F$-test compares the full Fourier model (Equation~\ref{eq:gam}) against an
intercept-only null model. The $p$-value quantifies whether the Fourier terms
jointly explain significant variance beyond the mean. Deviance explained (\%)
is reported as a measure of effect size.

\subsubsection{Confidence intervals}
Bootstrap 95\% confidence intervals are computed by resampling users with
replacement (200 bootstrap samples)~\citep{efron1993bootstrap}. For each
bootstrap sample, the full pipeline (normalisation and GAM fitting) is
repeated, and the 2.5th and 97.5th percentiles of the fitted values across
bootstrap samples define the confidence band.

\subsubsection{Model assumption checking}

The Fourier-basis GAM assumes that residuals are approximately normally
distributed and homoscedastic across cycle days, and that two harmonics provide
adequate flexibility to capture the true cycle pattern. For continuous
physiological outcomes these assumptions are generally reasonable; for outcomes
with non-normal distributions (e.g.\ binary or count symptom reports), $F$-test
$p$-values should be interpreted with appropriate caution. The choice of $K = 2$
harmonics balances flexibility against overfitting and performed well across the
outcomes examined here, though users should verify this is appropriate for their
data.

When applying the package to new datasets, users are encouraged to assess: (1)
normality of residuals via quantile--quantile plots, available as a diagnostic
output within the package; (2) homoscedasticity by inspecting residual plots
across cycle days; and (3) adequacy of the harmonic specification by comparing
model fit under alternative values of $K$ using the Akaike Information
Criterion, which the package reports automatically.

\subsection{Turning point detection and phase models}
\label{sec:turning}

\subsubsection{Turning point detection}
Peaks and troughs in the fitted GAM curve are identified by computing the first
derivative of the fitted function with respect to cycle day and locating sign
changes. A sign change from positive to negative indicates a peak (local
maximum); a sign change from negative to positive indicates a trough (local
minimum). Turning points are rounded to the nearest whole cycle day for
interpretability.

\subsubsection{Phase-specific linear models}
Between consecutive turning points, the package fits ordinary least-squares
linear models to the population-mean data, estimating the rate of change
(slope) in each phase. Standard errors are computed via the delta
method~\citep{oehlert1992delta}, and $p$-values test whether each slope differs
significantly from zero. These phase models provide clinically interpretable
summaries of the expected individual percentage change in outcome per day and
the direction of change within each cycle segment.

\subsection{Confounder analysis}
\label{sec:confounder}

Confounders are variables that may influence the outcome independently of the
menstrual cycle, potentially biasing estimates if not accounted for. The package
provides two approaches to confounder analysis:

\subsubsection{Confounder effect estimation}
To quantify the independent effect of each confounder on the outcome, the
package fits an augmented model that includes a linear confounder term alongside
the Fourier basis:
\begin{equation}
  y_i = \beta_0 + \sum_{k=1}^{K} \left[ \beta_{2k-1}
  \sin\!\left(\frac{2\pi k \cdot d_i}{28}\right) + \beta_{2k}
  \cos\!\left(\frac{2\pi k \cdot d_i}{28}\right) \right]
  + \gamma \cdot C_{i1} + \varepsilon_i
\end{equation}
where $C_{i1}$ is the confounder value for user $i$ and $\gamma$ is the
confounder coefficient, representing the change in outcome (in percentage points
from individual mean) per unit increase in the confounder, after accounting for
the cycle effect. A non-significant $\gamma$ indicates that the confounder does
not explain additional variance beyond the Fourier terms.

\subsubsection{Adjusted cycle effect model}
To visualise how the cycle effect changes after controlling for confounders, the
package can fit an adjusted GAM that includes multiple confounders
simultaneously:
\begin{equation}
  y_i = \beta_0 + \sum_{k=1}^{K} \left[ \beta_{2k-1}
  \sin\!\left(\frac{2\pi k \cdot d_i}{28}\right) + \beta_{2k}
  \cos\!\left(\frac{2\pi k \cdot d_i}{28}\right) \right]
  + \sum_{j=1}^{J} \gamma_j \cdot C_{ij} + \varepsilon_i
\end{equation}
where $C_{ij}$ represents the $j$-th confounder for observation $i$. Predictions
are generated with confounders held at their population means, yielding an
adjusted cycle curve that can be compared visually to the unadjusted curve.
Substantial divergence between curves indicates that confounders explain part of
the apparent cycle effect.

\subsection{Effect modifier analysis}
\label{sec:effectmod}

Effect modifiers are variables that change the strength or shape of the cycle
effect itself across subgroups. A confounder shifts the overall outcome level
but leaves the cycle pattern unchanged across groups; an effect modifier changes
the cycle pattern itself --- for example, producing a larger amplitude or
different timing in one group compared to another. Effect modifiers are therefore
examined by stratification rather than statistical adjustment.

\subsubsection{Stratified GAM analysis}
To assess effect modification, the package stratifies the data by levels of a
candidate modifier variable and fits separate Fourier-basis GAMs to each
stratum. For continuous modifiers, the package automatically creates groups
(Low: 0--25\% quartile; Medium: middle 50\%; High: top 25\%) with explicit
numeric boundaries displayed in the output. For each stratum $g$, the model is:
\begin{equation}
  y_{ig} = \beta_{0g} + \sum_{k=1}^{K} \left[ \beta_{(2k-1)g}
  \sin\!\left(\frac{2\pi k \cdot d_i}{28}\right) + \beta_{2kg}
  \cos\!\left(\frac{2\pi k \cdot d_i}{28}\right) \right] + \varepsilon_{ig}
\end{equation}
The fitted curves for each stratum are overlaid on a single plot, enabling
visual comparison of cycle patterns. Divergent curves (different amplitudes,
phase shifts, or turning point locations) suggest effect modification.

\subsubsection{Interpretation}
\label{sec:effectmod_interp}
The package reports for each stratum: sample size (users and observations),
cycle effect $p$-value, and deviance explained. A significant cycle effect in
one stratum but not another, or markedly different effect sizes across strata,
indicates that the modifier variable influences the cycle--outcome relationship.
This approach is exploratory; formal interaction testing would require fitting a
single model with cycle-by-modifier interaction terms.

It should be noted that the distinction between confounders and effect modifiers
is not always stable across sample sizes. A variable that appears to modify the
shape or amplitude of the cycle effect in a small sample may reveal confounding
properties as sample size increases, as greater statistical power exposes
previously undetectable main effects on the outcome. Where a candidate variable
is plausibly both a confounder and an effect modifier, researchers are encouraged
to apply both the adjustment approach (Section~\ref{sec:confounder}) and the
stratification approach (Section~\ref{sec:effectmod}) and compare results.
Substantial divergence between adjusted and stratified estimates would suggest
that the variable operates through both pathways and warrants further
investigation in larger samples.

\subsection{Causal inference framework}
\label{sec:causal}

The confounder and effect modifier analyses in \texttt{mcanalysis} are designed
to support causal inference about menstrual cycle effects. While observational
data cannot definitively establish causality, the package provides tools to
strengthen causal claims by addressing key threats to internal validity.

\subsubsection{Confounder adjustment for causal identification}
The menstrual cycle is driven by endogenous hormonal fluctuations and cannot be
randomised experimentally. However, within-person normalisation
(Section~\ref{sec:normalization}) eliminates time-invariant confounding by
comparing each individual to their own baseline, analogous to a fixed-effects
design~\citep{allison2009fixed}. This approach controls for stable individual
characteristics (e.g., genetics, chronic health conditions, baseline physiology)
that might otherwise confound the cycle--outcome relationship.

Time-varying confounders --- variables that change day-to-day and may correlate
with both cycle phase and the outcome --- require explicit adjustment. The
confounder analysis module (Section~\ref{sec:confounder}) allows researchers to
test and control for potential time-varying confounders such as physical
activity, sleep, or external stressors. If the cycle effect estimate remains
stable after confounder adjustment, this strengthens the causal interpretation
by ruling out measured confounding pathways.

\subsubsection{Effect modification for heterogeneous causal effects}
Causal effects may vary across individuals or contexts, a phenomenon known as
effect heterogeneity or effect
modification~\citep{vanderweele2007four}. The effect modifier analysis
(Section~\ref{sec:effectmod}) enables researchers to examine whether the cycle
effect on the outcome differs by individual characteristics (e.g.\ age, BMI,
disease severity). Identifying effect modifiers has practical implications: it
reveals for whom interventions timed to cycle phase may be most beneficial,
supporting personalised approaches to women's health.

\subsubsection{Limitations of causal inference}
Despite these tools, causal claims from \texttt{mcanalysis} remain limited by
unmeasured confounding. Without direct physiological measurements such as
hormone assays, the biological pathway from cycle phase to outcome cannot be
verified. The package quantifies associations that are consistent with a causal
interpretation when confounders are adequately controlled, but users should
interpret results within the constraints of observational data and consider
triangulation with experimental or Mendelian randomisation designs where
feasible~\citep{lawlor2016triangulation}.

\subsection{Visualisation}
\label{sec:visualization}

The package generates publication-ready plots that integrate multiple layers of
information:
\begin{itemize}
  \item The fitted GAM curve with 95\% bootstrap confidence interval (shaded
    band)
  \item Daily population-mean data points (optional raw data overlay)
  \item Turning point markers (peaks and troughs) with annotations
  \item Phase-specific linear model overlays with slope estimates (in \%
    expected daily change) and significance
  \item Phase shading from derived turning points
  \item Significance annotations ($p$-value, deviance explained)
\end{itemize}
For confounder analyses, the package produces forest plots displaying
coefficients with 95\% confidence intervals for each tested confounder. All
visualisations are customisable (colours, labels, figure dimensions) and can be
saved to file.

\section{Results}

\subsection{Dataset characteristics}

Table~\ref{tab:dataset} summarises the Juli dataset characteristics. Of 19,977
total app users, 2,816 had recorded period dates and were included in the
menstrual cycle analysis.

\begin{table}[H]
\centering
\caption{Juli dataset characteristics.}
\label{tab:dataset}
\begin{tabular}{ll}
\toprule
\textbf{Characteristic} & \textbf{Value} \\
\midrule
Population                  & Adults using Juli App \\
Users with period data      & 2,816 \\
Total observations          & 188,816 \\
Number of outcomes analysed & 15 \\
Outcome domains             & Physiology, Sleep, Mental Health, Activity \\
Confounder variables        & Age, mean steps, mean sleep \\
Significant outcomes        & 9/15 \\
\bottomrule
\end{tabular}
\end{table}

\subsection{Menstrual cycle effects}

Of the outcomes analysed, 9 showed evidence of association with the menstrual
cycle at $p < 0.05$ (Table~\ref{tab:mc_results}).

\subsubsection{Physiological outcomes}

\begin{figure}[H]
\centering
\begin{subfigure}[b]{0.48\textwidth}
    \centering
    \includegraphics[width=\textwidth]{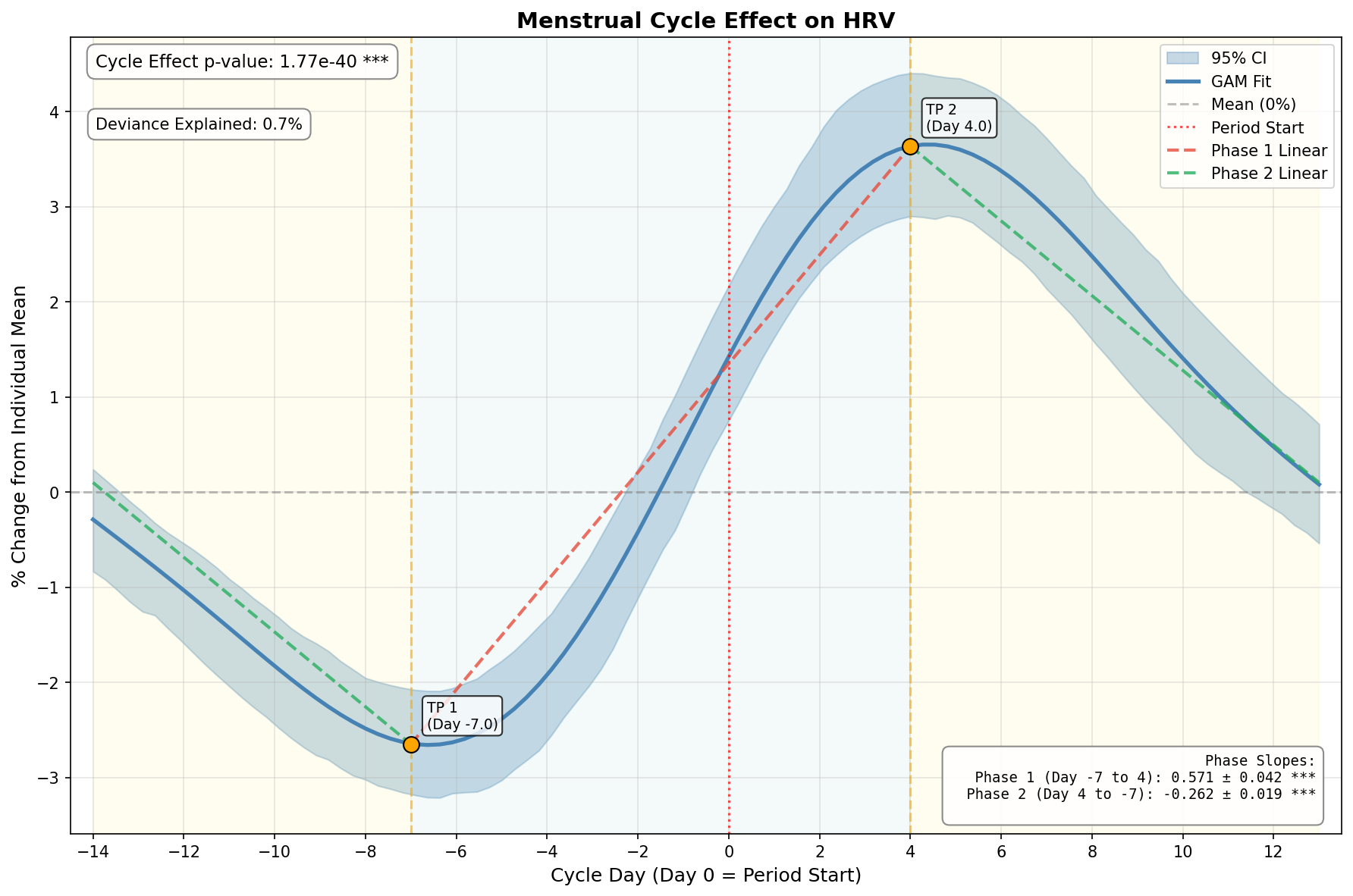}
    \caption{Heart rate variability}
\end{subfigure}
\hfill
\begin{subfigure}[b]{0.48\textwidth}
    \centering
    \includegraphics[width=\textwidth]{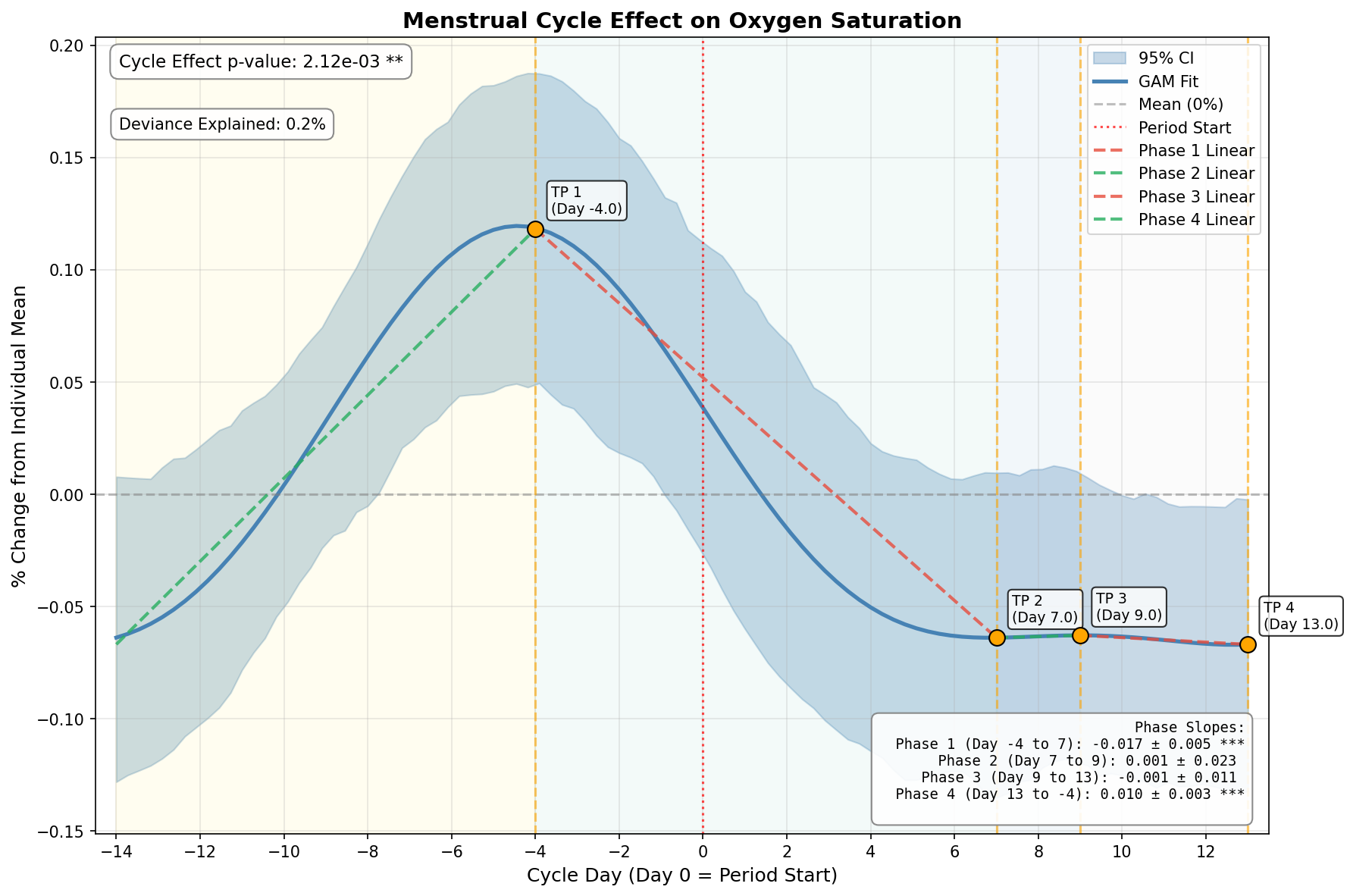}
    \caption{Oxygen saturation}
\end{subfigure}
\caption{Menstrual cycle effects on physiological outcomes (Juli dataset).
\textbf{(A)}~HRV shows evidence of a cycle effect ($p < 0.001$, deviance
explained $= 0.66\%$, $N = 325$ users, 25,280 observations), with a trough at
Day~$-7$ (pre-menstrual) and peak at Day~$+4$ (post-menstrual). Phase-specific
linear models (dashed lines) indicate a rise of $+0.65 \pm 0.05$\,\%/day from
trough to peak and a decline of $-0.26 \pm 0.02$\,\%/day from peak to trough.
\textbf{(B)}~Oxygen saturation also shows evidence of a cycle effect
($p = 0.002$, deviance explained $= 0.66\%$, $N = 132$ users, 10,203
observations), with a complex pattern characterised by four turning points at
Days~$-4$, 7, 9, and 13. This is a novel finding and warrants cautious
interpretation pending replication. The fitted Fourier-basis GAM curve (solid
line) is shown with 95\% bootstrap confidence interval (shaded band).
Day~0 = menstruation onset.}
\label{fig:physiological}
\end{figure}

HRV showed the strongest evidence of a cycle effect ($p < 0.001$, $n = 325$
users, 25,280 observations), with a trough at Day~$-7$ (pre-menstrual) and peak
at Day~$+4$ (post-menstrual), consistent with progesterone-driven reductions in
parasympathetic tone in the luteal phase
(Figure~\ref{fig:physiological}A). Oxygen saturation also showed evidence of a
cycle effect ($p = 0.002$, $n = 132$ users, 10,203 observations), with a more
complex pattern characterised by four turning points at Days~$-4$, 7, 9, and 13
(Figure~\ref{fig:physiological}B). This is a novel finding and warrants cautious
interpretation pending replication.

\subsubsection{Sleep}

Time asleep showed evidence of a cycle effect ($p = 0.003$, $n = 495$ users,
36,295 observations), with turning points at Days~$-8$, 0, 7, and 13,
suggesting a complex pattern of sleep duration variation across the cycle
(Figure~\ref{fig:sleep_activity}A). This finding is consistent with prior
evidence of menstrual cycle effects on sleep architecture.

\subsubsection{Pain and symptom outcomes}

\begin{figure}[H]
\centering
\begin{subfigure}[b]{0.48\textwidth}
    \centering
    \includegraphics[width=\textwidth]{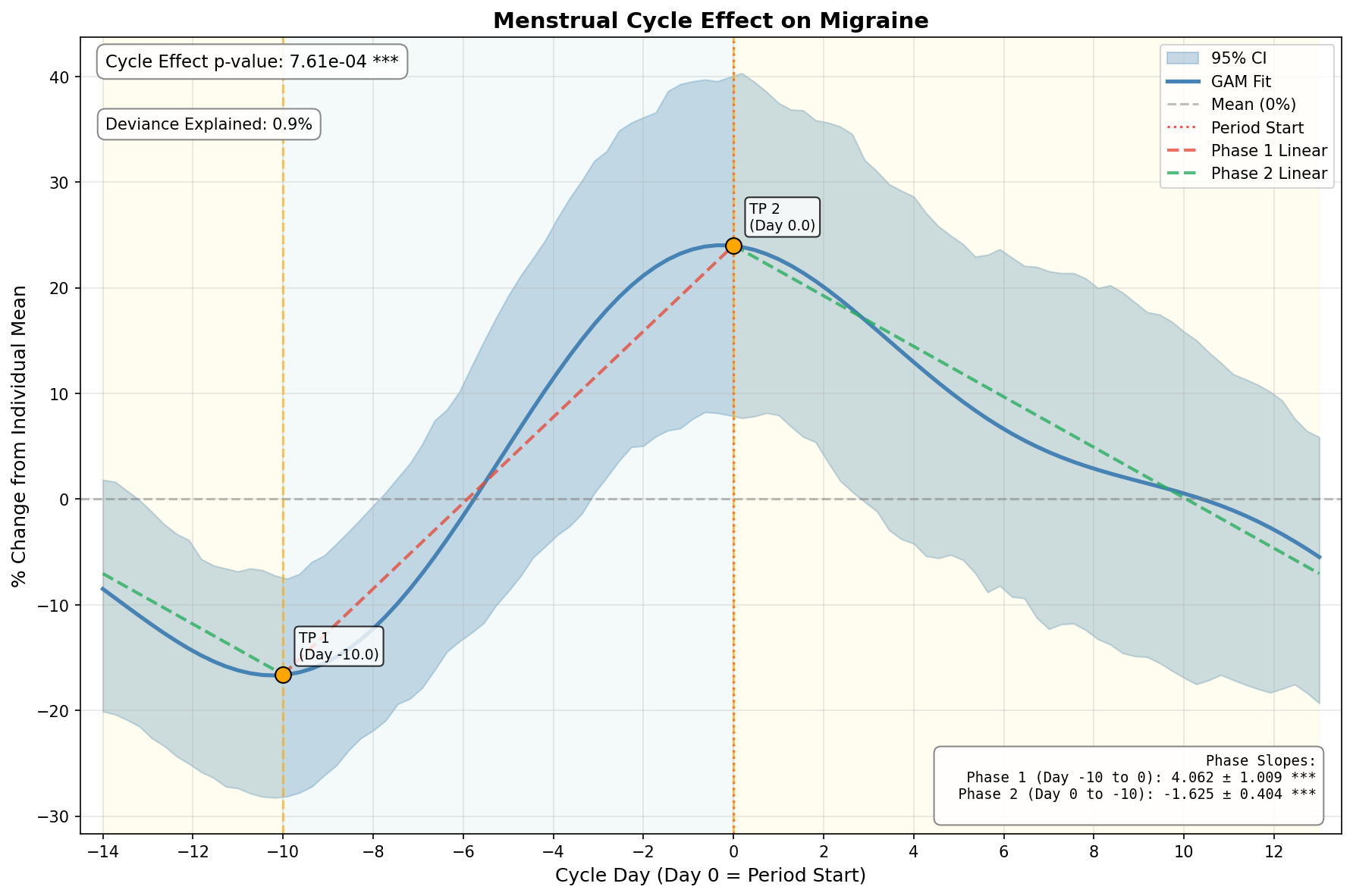}
    \caption{Migraine}
\end{subfigure}
\hfill
\begin{subfigure}[b]{0.48\textwidth}
    \centering
    \includegraphics[width=\textwidth]{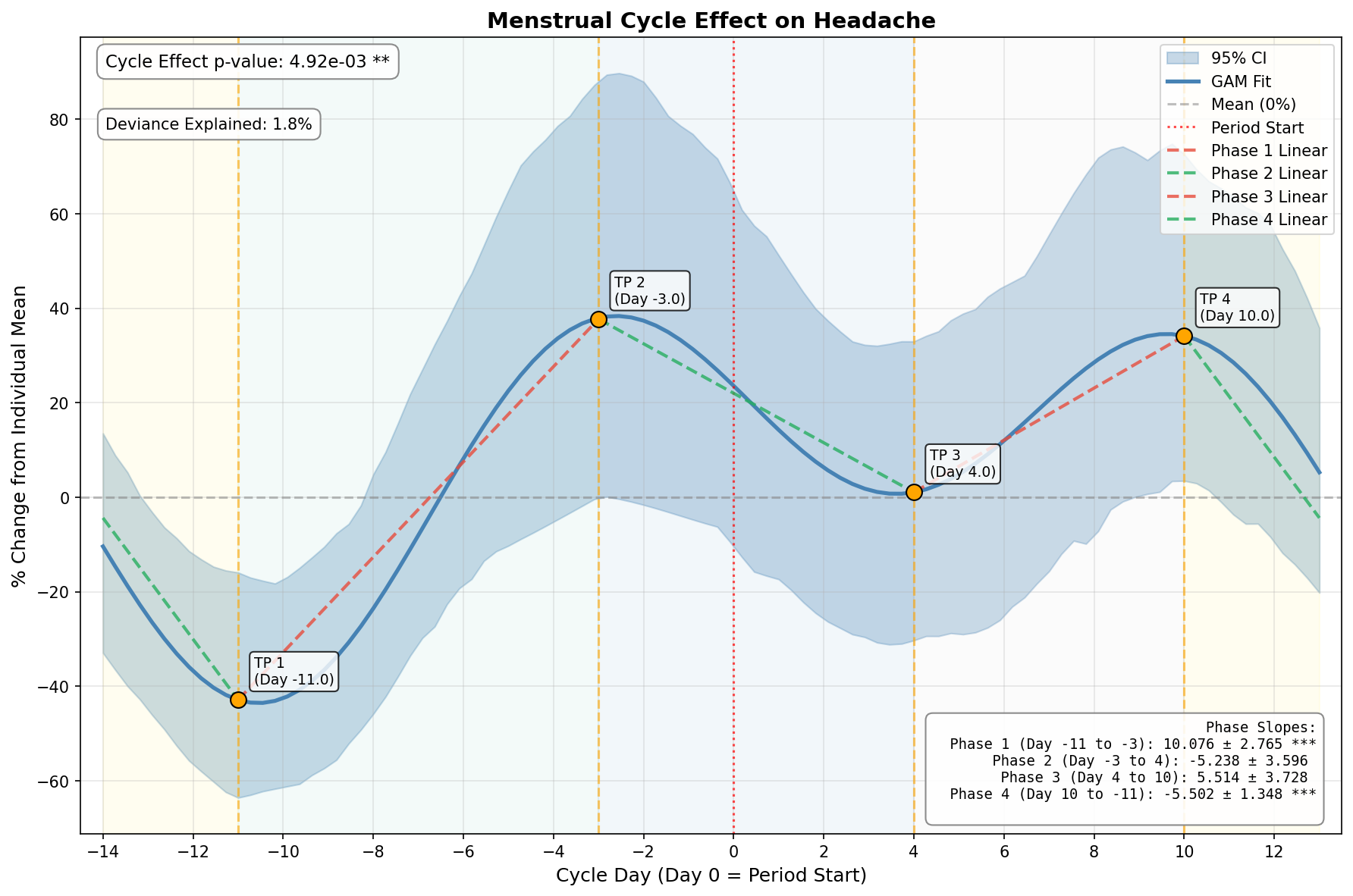}
    \caption{Headache}
\end{subfigure}
\caption{Menstrual cycle effects on pain and symptom outcomes (Juli dataset).
\textbf{(A)}~Migraine shows evidence of a cycle effect ($p < 0.001$, deviance
explained $= 0.93\%$, $N = 34$ users, 2,028 observations), with turning points
at Day~$-10$ and Day~0, consistent with the well-documented pattern of menstrual
migraine linked to oestrogen withdrawal around menstruation
onset~\citep{macgregor2004incidence, somerville1972influence}.
\textbf{(B)}~Headache also shows evidence of a cycle effect ($p = 0.005$,
deviance explained $= 1.51\%$, $N = 17$ users, 803 observations), with turning
points at Days~$-11$, $-3$, 4, and 10. The fitted Fourier-basis GAM curve
(solid line) is shown with 95\% bootstrap confidence interval (shaded band).
Day~0 = menstruation onset.}
\label{fig:symptoms}
\end{figure}

Migraine showed a highly significant cycle effect ($p < 0.001$, $n = 34$ users,
2,028 observations), with turning points at Day~$-10$ and Day~0, consistent
with the well-documented pattern of menstrual migraine linked to oestrogen
withdrawal around menstruation
onset~\citep{macgregor2004incidence, somerville1972influence}
(Figure~\ref{fig:symptoms}A). Headache also showed evidence of a cycle effect
($p = 0.005$, $n = 17$ users, 803 observations), with four turning points at
Days~$-11$, $-3$, 4, and 10 (Figure~\ref{fig:symptoms}B). General pain
($p = 0.524$) and asthma/shortness of breath ($p = 0.508$) did not show
evidence of cycle effects.

\subsubsection{Mood and mental health outcomes}

\begin{figure}[H]
\centering
\begin{subfigure}[b]{0.48\textwidth}
    \centering
    \includegraphics[width=\textwidth]{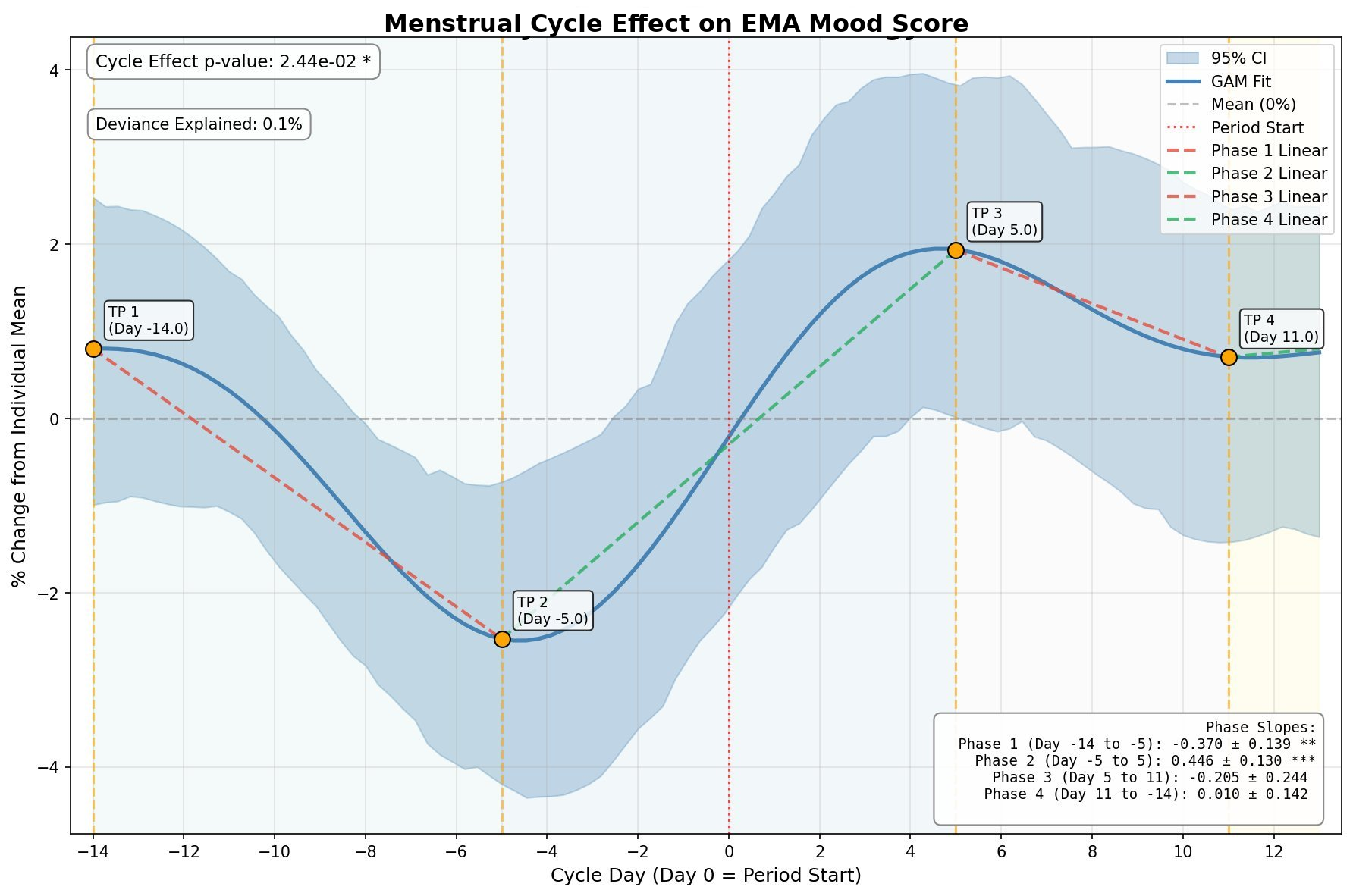}
    \caption{EMA mood}
\end{subfigure}
\hfill
\begin{subfigure}[b]{0.48\textwidth}
    \centering
    \includegraphics[width=\textwidth]{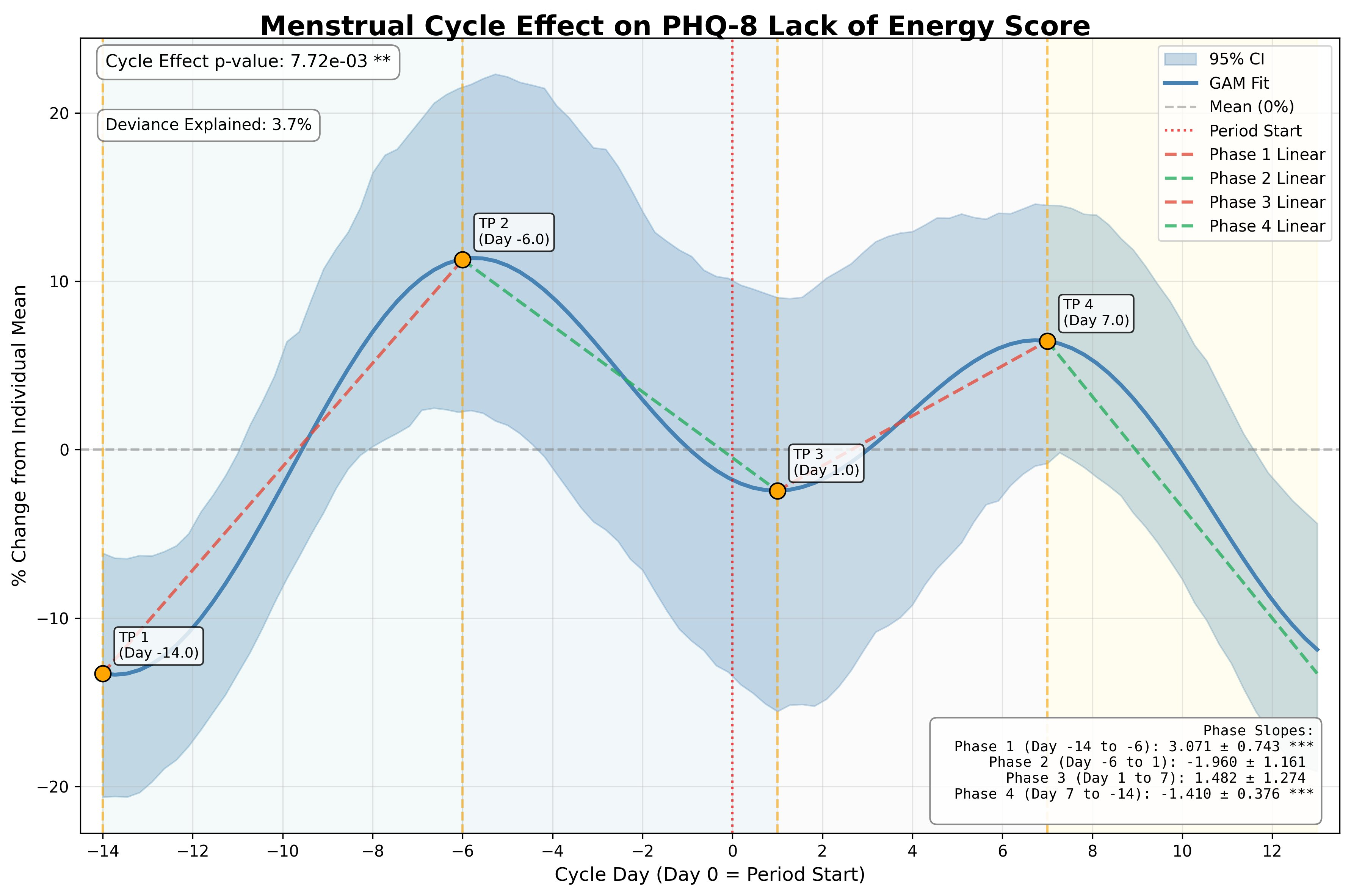}
    \caption{PHQ-8 lack of energy}
\end{subfigure}

\vspace{0.5em}

\begin{subfigure}[b]{0.48\textwidth}
    \centering
    \includegraphics[width=\textwidth]{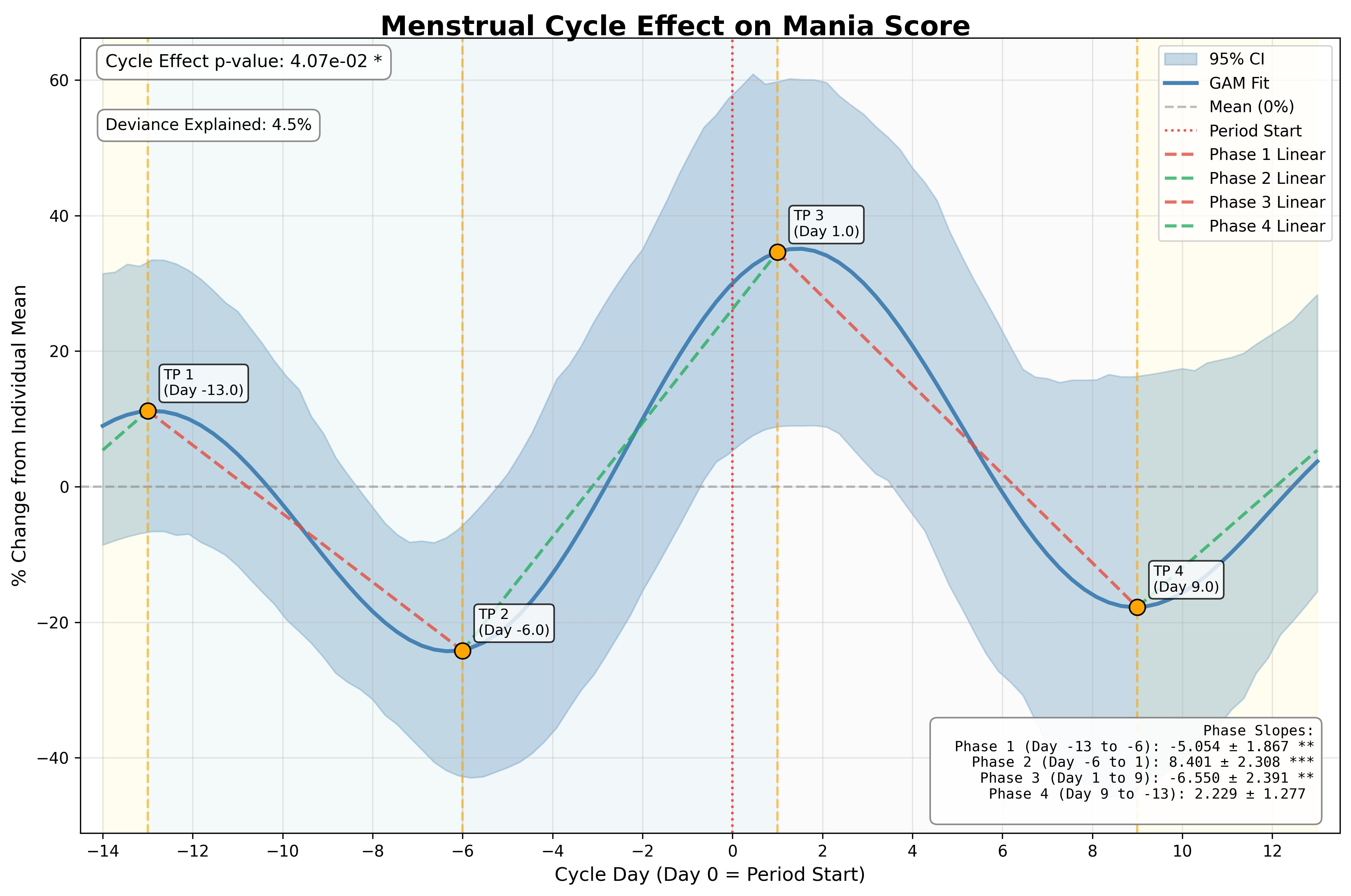}
    \caption{Mania}
\end{subfigure}
\caption{Menstrual cycle effects on mental health outcomes (Juli dataset).
\textbf{(A)}~EMA mood ratings show evidence of a cycle effect ($p = 0.024$,
deviance explained $= 0.02\%$, $N = 194$ users, 9,431 observations), with a
trough at Day~$-5$ and peak at Day~$+5$, mirroring the HRV pattern.
\textbf{(B)}~PHQ-8 lack of energy shows evidence of a cycle effect ($p = 0.008$,
deviance explained $= 3.73\%$, $N = 18$ users, 359 observations), with turning
points at Days~$-14$, $-6$, 1, and 7. \textbf{(C)}~Mania scores show evidence
of a cycle effect ($p = 0.041$, deviance explained $= 4.50\%$, $N = 11$ users,
216 observations), with turning points at Days~$-13$, $-6$, 1, and 9. Note that
EMA mood and EMA energy ($p = 0.156$) show a notable dissociation despite being
collected via the same instrument, suggesting distinct cycle-related profiles for
mood and energy. PHQ-8 and mania findings should be interpreted with caution
given the small user samples; replication in larger cohorts is needed. The fitted
Fourier-basis GAM curve (solid line) is shown with 95\% bootstrap confidence
interval (shaded band). Day~0 = menstruation onset.}
\label{fig:mental_health}
\end{figure}

EMA mood ratings showed evidence of a cycle effect ($p = 0.024$, $n = 194$
users, 9,431 observations), with a trough at Day~$-5$ and peak at Day~$+5$
(Figure~\ref{fig:mental_health}A). This contrasts with EMA energy ratings, which
did not show evidence of a cycle effect ($p = 0.156$), suggesting that mood and
energy, though collected via the same EMA instrument, have distinct
cycle-related profiles. Lack of energy as assessed by the PHQ-8 showed evidence
of a cycle effect ($p = 0.008$, $n = 18$ users, 359 observations), with four
turning points at Days~$-14$, $-6$, 1, and 7
(Figure~\ref{fig:mental_health}B). Mania scores also showed evidence of a cycle
effect ($p = 0.041$, $n = 11$ users, 216 observations), with turning points at
Days~$-13$, $-6$, 1, and 9 (Figure~\ref{fig:mental_health}C). Both PHQ-8 and
mania findings should be interpreted with caution given the small user samples,
which limit statistical power and increase the risk of inflated effect
estimates~\citep{button2013power}. It is also worth noting that under a
Benjamini--Hochberg false discovery rate (FDR) correction for multiple
testing~\citep{benjamini1995controlling} applied across all 15 outcomes, mania
($p = 0.041$) would no longer meet the significance threshold, while all other
eight significant outcomes would retain significance. The mania finding should
therefore be treated as exploratory pending replication in a larger sample.

\subsubsection{Activity outcomes}

\begin{figure}[H]
\centering
\begin{subfigure}[b]{0.48\textwidth}
    \centering
    \includegraphics[width=\textwidth]{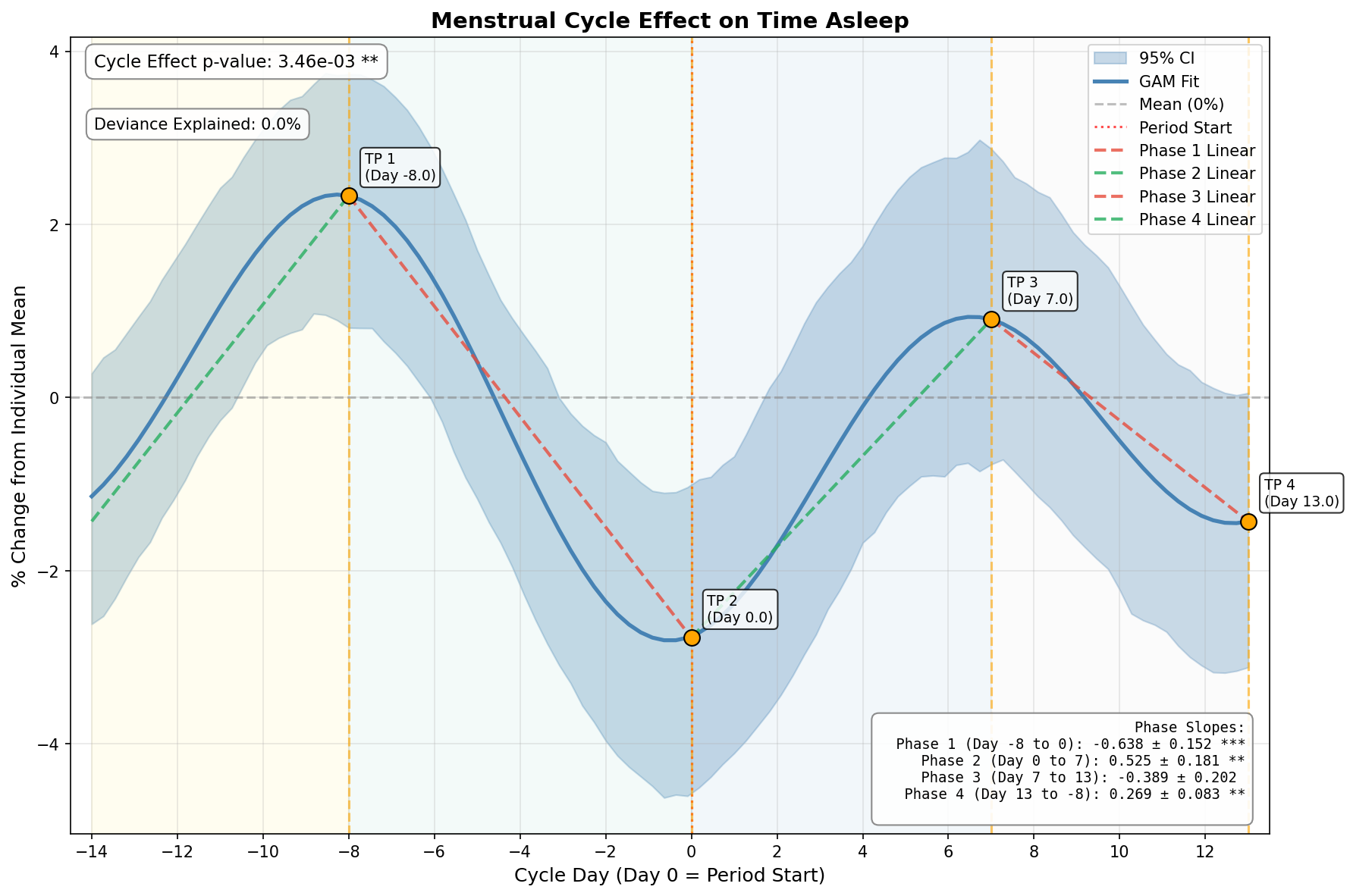}
    \caption{Time asleep}
\end{subfigure}
\hfill
\begin{subfigure}[b]{0.48\textwidth}
    \centering
    \includegraphics[width=\textwidth]{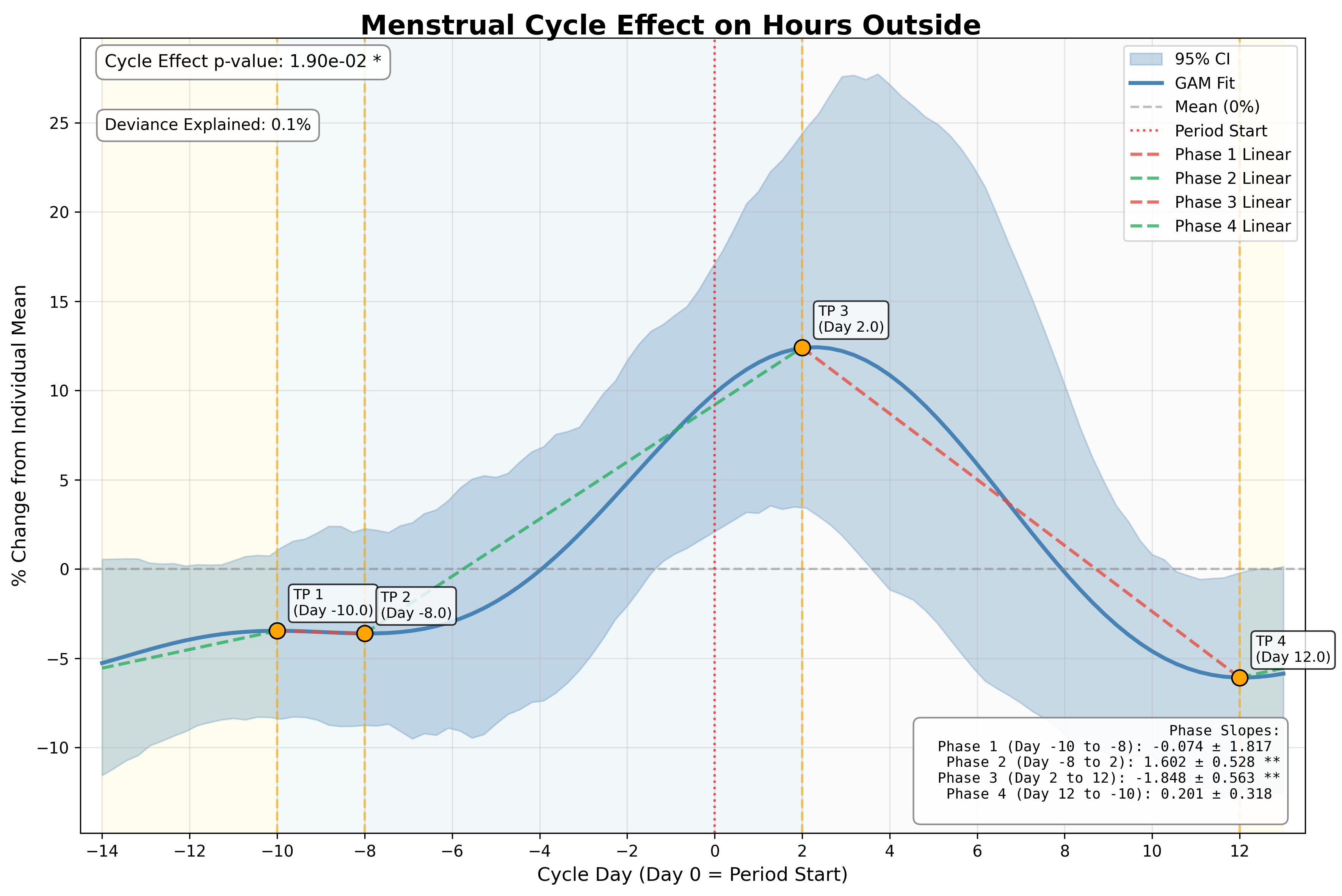}
    \caption{Hours outside}
\end{subfigure}
\caption{Menstrual cycle effects on sleep and activity outcomes (Juli dataset).
\textbf{(A)}~Time asleep shows evidence of a cycle effect ($p = 0.003$, deviance
explained $= 0.11\%$, $N = 495$ users, 36,295 observations), with turning points
at Days~$-8$, 0, 7, and 13. \textbf{(B)}~Hours outside shows evidence of a
cycle effect ($p = 0.019$, deviance explained $= 0.15\%$, $N = 157$ users,
7,970 observations), with turning points at Days~$-10$, $-8$, 2, and 12. Both
outcomes show small deviance explained values, indicating that the menstrual
cycle accounts for a modest fraction of day-to-day variability in these measures.
The fitted Fourier-basis GAM curve (solid line) is shown with 95\% bootstrap
confidence interval (shaded band). Day~0 = menstruation onset.}
\label{fig:sleep_activity}
\end{figure}

Hours outside showed evidence of a cycle effect ($p = 0.019$, $n = 157$ users,
7,970 observations), with turning points at Days~$-10$, $-8$, 2, and 12
(Figure~\ref{fig:sleep_activity}B). Steps ($p = 0.315$), active energy
($p = 0.396$), and basal energy ($p = 0.821$) did not show evidence of cycle
effects, suggesting that structured or passive activity metrics are less
sensitive to cycle-related variation than time spent outdoors.

\begin{table}[h]
\centering
\caption{Menstrual cycle effects on health outcomes.}
\label{tab:mc_results}
\begin{threeparttable}
\begin{tabular}{llrrll}
\toprule
\textbf{Category} & \textbf{Outcome} & \textbf{N users} & \textbf{N obs}
  & \textbf{$p$-value} & \textbf{Turning points} \\
\midrule
\multirow{2}{*}{\textit{Physiological}}
  & HRV               & 325 & 25,280 & $<0.001$*** & Day $-7$, 4 \\
  & Oxygen Saturation & 132 & 10,203 & 0.002**     & Day $-4$, 7, 9, 13 \\
\midrule
\textit{Sleep}
  & Time Asleep       & 495 & 36,295 & 0.003**     & Day $-8$, 0, 7, 13 \\
\midrule
\multirow{4}{*}{\textit{Pain/Symptoms}}
  & Migraine          &  34 &  2,028 & $<0.001$*** & Day $-10$, 0 \\
  & Headache          &  17 &    803 & 0.005**     & Day $-11$, $-3$, 4, 10 \\
  & Pain              &  47 &  2,089 & 0.524       & --- \\
  & Asthma (SOB)      &  48 &  3,145 & 0.508       & --- \\
\midrule
\multirow{4}{*}{\textit{Mood/Mental Health}}
  & EMA Mood              & 194 &  9,432 & 0.024*  & Day $-5$, 5 \\
  & EMA Energy            & 194 &  9,432 & 0.156   & --- \\
  & Lack of Energy (PHQ-8) &  18 &    359 & 0.008** & Day $-14$, $-6$, 1, 7 \\
  & Mania                 &  11 &    216 & 0.041*  & Day $-13$, $-6$, 1, 9 \\
\midrule
\multirow{4}{*}{\textit{Activity}}
  & Hours Outside &  157 &  7,970 & 0.019*  & Day $-10$, $-8$, 2, 12 \\
  & Steps         &  716 & 56,635 & 0.315   & --- \\
  & Active Energy &  122 &  6,843 & 0.396   & --- \\
  & Basal Energy  &  486 & 39,707 & 0.821   & --- \\
\bottomrule
\end{tabular}
\begin{tablenotes}
\small
\item $^{*}p < 0.05$, $^{**}p < 0.01$, $^{***}p < 0.001$
\end{tablenotes}
\end{threeparttable}
\end{table}

\subsection{Non-significant outcomes}

Six of 15 outcomes did not show evidence of cycle effects at $p < 0.05$:
general pain ($p = 0.524$), asthma/shortness of breath ($p = 0.508$), EMA
energy ratings ($p = 0.156$), steps ($p = 0.315$), active energy ($p = 0.396$),
and basal energy ($p = 0.821$). The non-significance of EMA energy is
particularly noteworthy given that EMA mood, collected via the same instrument
from the same users, showed evidence of a cycle effect ($p = 0.024$). This
dissociation suggests that mood and energy, though closely related constructs,
have distinct cycle-related profiles and should not be treated as
interchangeable outcomes in menstrual cycle research. The non-significance of
structured activity metrics (steps, active energy, basal energy) alongside the
significance of hours outside is consistent with the interpretation that
voluntary behaviour may be more sensitive to cycle-related variation than
passive physiological measures of energy expenditure.

\subsection{Confounder analysis}

Confounder analyses were performed for the primary outcomes
(Table~\ref{tab:confounders}). None of the tested confounders showed evidence
of association with cycle-normalised outcomes. The near-unity $p$-values
observed for several confounders are expected given the within-person
normalisation applied in preprocessing: by expressing each observation as a
percentage of the individual's own mean, stable between-person characteristics
such as age and baseline activity levels are largely absorbed into the
normalisation, leaving little residual variance for these confounders to explain
beyond the Fourier terms. This is consistent with the fixed-effects
interpretation of the normalisation procedure~\citep{allison2009fixed}, and
indicates that the observed cycle effects are not materially confounded by age,
activity levels, or sleep duration.

\begin{table}[H]
\centering
\caption{Confounder analysis results. No confounder showed evidence of
association with cycle-normalised outcomes for the primary outcomes HRV and
mood.}
\label{tab:confounders}
\small
\begin{tabular}{llrrl}
\toprule
\textbf{Outcome} & \textbf{Confounder} & \textbf{Coefficient}
  & \textbf{95\% CI} & \textbf{$p$-value} \\
\midrule
\multirow{3}{*}{HRV}
  & Age        & $0.0007$  & [$-0.036$,  $0.038$]   & 0.97 \\
  & Mean steps & $-0.0000$ & [$-0.0001$, $0.0001$]  & 0.94 \\
  & Mean sleep & $0.0010$  & [$-0.093$,  $0.095$]   & 0.98 \\
\midrule
\multirow{3}{*}{Mood}
  & Age        & $-0.0000$ & [$-0.045$,  $0.045$]   & 1.00 \\
  & Mean steps & $-0.0000$ & [$-0.0001$, $0.0001$]  & 0.98 \\
  & Mean sleep & $0.0003$  & [$-0.121$,  $0.122$]   & 1.00 \\
\bottomrule
\end{tabular}
\end{table}

\section{Discussion}

This study presents \texttt{mcanalysis}, an open-source, dual-language package
for full menstrual cycle analysis from preprocessing through statistical
modelling to publication-ready visualisation, demonstrated on a large dataset of
individuals using the Juli chronic health management application. Our findings
demonstrate that physiological, sleep, symptom, mood, and activity outcomes all
show detectable menstrual cycle effects, with implications for understanding how
the menstrual cycle relates to health across multiple domains. Nine of the
outcomes analysed showed evidence of cycle effects, spanning five distinct
outcome categories, demonstrating the breadth of the package's utility across
condition types and measurement modalities.

This paper reproduces established findings to demonstrate the validity of the
dataset and modelling methods, and presents new findings on variables associated
with the menstrual cycle.

\subsection{Physiological cycle effects}

HRV showed a consistent cycle effect, with lower HRV pre-menstrually
(Day~$-7$) and higher HRV post-menstrually (Day~$+4$). This pattern is
consistent with the known sympathoexcitatory effects of progesterone, which
peaks pre-menstrually and is associated with reduced parasympathetic
tone~\citep{grant2020menstrual, brar2015effect, yildirir2002effects,
leicht2003heart}. The consistency of this finding with prior literature supports
the validity of the \texttt{mcanalysis} pipeline.

The magnitude of the HRV effect (deviance explained $= 0.66\%$) is consistent
with the expectation that cycle effects represent a modest but detectable source
of physiological variation, superimposed on larger sources of day-to-day
variability. The strong evidence of association reflected in the $p$-value is in
part a function of the large sample size ($n = 325$ users, 25,280 observations);
the effect size indicates that the menstrual cycle accounts for a small but
reliably detectable fraction of HRV variance in daily life.

Oxygen saturation also showed evidence of a cycle effect ($p = 0.002$,
$n = 132$ users, 10,203 observations), with a complex pattern characterised by
four turning points at Days~$-4$, 7, 9, and 13. This is a novel finding. Prior
literature suggests that respiratory function may vary across the
cycle~\citep{baker2007circadian}, which could contribute to observed variation
in peripheral oxygen saturation, though the mechanisms underlying this
association remain to be established. The multi-peak pattern warrants cautious
interpretation, and replication in larger samples is needed before further
conclusions are drawn.

\subsection{Sleep}

Time asleep showed evidence of a cycle effect ($p = 0.003$, $n = 495$ users,
36,295 observations), with turning points at Days~$-8$, 0, 7, and 13,
suggesting systematic variation in sleep duration across the cycle. This is
consistent with prior evidence of menstrual cycle effects on sleep architecture,
including reduced slow-wave sleep and altered sleep efficiency in the luteal
phase~\citep{baker2007circadian, dezambotti2015menstrual}. The relatively large
sample size for this outcome strengthens confidence in the result and suggests
that sleep duration is a detectable marker of cycle-related variation via
consumer wearables.

\subsection{Mental health outcomes}

EMA mood ratings showed evidence of a cycle effect ($p = 0.024$, $n = 194$
users, 9,431 observations), with a trough at Day~$-5$ and peak at Day~$+5$,
mirroring the pattern observed for HRV. This finding is notable given prior
meta-analytic evidence suggesting that premenstrual mood changes are small when
measured
prospectively~\citep{romans2012mood, schmalenberger2021menstrual,
toffoletto2014emotional}. The detection of an association here may reflect the
scale of the dataset, the sensitivity of daily EMA measurement, or
characteristics specific to this clinical population. The effect size (deviance
explained $= 0.02\%$) is small, and caution is warranted in interpreting
clinical significance from a statistically significant but modest association.

A notable dissociation emerged between EMA mood and EMA energy ratings: despite
being collected via the same instrument and from the same users ($n = 194$), EMA
mood showed evidence of association ($p = 0.024$) while EMA energy did not
($p = 0.156$). This suggests that mood and energy, though often treated as
closely related constructs, have distinct cycle-related profiles. A
complementary dissociation was observed between the PHQ-8 lack-of-energy item,
which showed evidence of a cycle effect ($p = 0.008$), and the EMA energy
ratings, which did not. This divergence may reflect differences in the timescale
of measurement (retrospective clinical item vs.\ momentary self-report) or in
the construct being captured. Taken together, these dissociations highlight the
importance of outcome measure selection in menstrual cycle research, and suggest
that clinical questionnaire items and momentary ecological assessments should not
be treated as interchangeable when studying cycle effects.

Mania scores also showed evidence of a cycle effect ($p = 0.041$, $n = 11$
users, 216 observations). Both the PHQ-8 and mania findings must be interpreted
cautiously given the small user samples, which limit statistical power and
increase the risk of inflated effect estimates~\citep{button2013power}. These
results should be treated as hypothesis-generating, and replication in larger
clinical cohorts is needed before conclusions are drawn about cycle effects on
clinical mental health symptomatology.

\subsection{Symptom outcomes}

The evidence of cycle effects for migraine and headache aligns with clinical
observations of menstrual migraine, a well-documented phenomenon linked to
oestrogen withdrawal around
menstruation~\citep{macgregor2004incidence, somerville1972influence}. The
turning point at Day~0 for migraine is consistent with the typical timing of
menstrual migraine onset. In contrast, general pain ($p = 0.524$) and
asthma/shortness of breath ($p = 0.508$) did not show evidence of cycle
effects, suggesting that not all symptom outcomes are equally sensitive to
cycle-related variation and that the package's significance testing
appropriately discriminates between outcomes.

\subsection{Activity outcomes}

Hours outside showed evidence of a cycle effect ($p = 0.019$, $n = 157$ users,
7,970 observations), while steps, active energy, and basal energy did not. This
pattern suggests that voluntary outdoor behaviour may vary across the cycle in
ways that structured activity metrics do not capture, potentially reflecting
changes in motivation, fatigue, or social behaviour. Consistent with this
interpretation, basal energy as a passive metabolic measure would not be
expected to show cycle variation of the magnitude detectable in this dataset,
and its non-significance ($p = 0.821$) aligns with that expectation.

\subsection{Package contributions}

The \texttt{mcanalysis} package addresses several practical challenges in
menstrual cycle research. At its core, it provides a \emph{complete
preprocessing pipeline}, from raw period dates through cycle identification,
quality filtering, and normalisation, that standardises data preparation steps
which are often performed ad hoc. Building on this, the Fourier-basis GAM
ensures \emph{cyclic continuity}, avoiding the artificial discontinuity at cycle
boundaries that plagues standard regression approaches. The \emph{integrated
visualisation system} further produces publication-ready figures that layer GAM
curves, confidence intervals, turning points, and phase models in a single plot,
facilitating communication of complex cycle patterns to clinicians and
non-specialist audiences. Alongside this, the \emph{dual R/Python
implementation} ensures accessibility across the two dominant languages in
biomedical data science, with numerically identical results guaranteeing
cross-platform reproducibility. For users without programming expertise, the
\emph{web-based interface} enables cycle analyses to be conducted entirely
through a point-and-click environment. Beyond immediate analysis needs, the
\emph{cyclic structure} of the fitted models makes outputs directly integrable
into predictive models and personalised intervention frameworks such as dynamic
treatment regimes.

\subsection{Implications for research and practice}

These findings have several implications for how menstrual cycle effects can be
incorporated into personalised health monitoring and intervention. For
researchers, the \texttt{mcanalysis} package provides a standardised approach to
quantifying cycle-related variation across a broad range of outcomes,
facilitating cross-study comparison and enabling the field to build a cumulative
evidence base for which outcomes vary with cycle phase and by how much. For
digital health practitioners and app developers, the results demonstrate that
outcomes spanning physiology, sleep, symptoms, mood, and behaviour all show
detectable cycle-related variation, meaning that recommendations and
interpretations based on these outcomes --- whether for activity targets, sleep
norms, or symptom tracking --- may benefit from cycle-aware adjustment rather
than treating all days as equivalent.

More concretely, the cyclic structure of the fitted models provides a principled
basis for incorporating expected cycle-phase variation into personalised
recommendation systems. Rather than flagging a pre-menstrual dip in HRV or mood
as an anomaly, a cycle-aware system can contextualise it as an expected pattern
for that individual at that phase, and adjust targets or alerts accordingly.
This is particularly relevant for dynamic treatment regimes and just-in-time
adaptive interventions, where the timing of recommendations can be optimised to
account for predictable within-person variation. The finding that multiple
outcome domains show evidence of cycle effects suggests that such adjustments
are not limited to reproductive health applications but extend to general
wellness, mental health, and chronic condition management tools that serve
menstruating users.

\subsection{Limitations}

Several limitations should be noted. This is an observational analysis of
routinely collected data; causal inference about underlying biological
mechanisms is not possible without direct physiological measurements such as
hormone assays. Relatedly, period dates are self-reported and may contain recall
errors, which would attenuate true cycle
effects~\citep{symul2019assessment, fehring2006variability}. Ovulation was not
confirmed (e.g., via LH testing), so turning points are statistical estimates
rather than biologically validated phase boundaries. The Juli dataset also
comprises individuals with a range of chronic health conditions, which may
influence both physiological and self-reported cycle patterns; findings may not
generalise to the general population. While deviance explained provides a
standardised effect size, the relatively small values for most outcomes indicate
that the menstrual cycle accounts for a modest fraction of day-to-day
variability, and associations should not be interpreted as clinically large on
the basis of their $p$-values alone. The small user samples for mental health
outcomes ($n = 11$--$18$) further limit confidence in those specific findings,
and the risk of inflated effect estimates in small samples should be
noted~\citep{button2013power}. Additionally, under a Benjamini--Hochberg FDR
correction across all 15 outcomes, mania would no longer retain significance,
and results for this outcome should be interpreted accordingly. Looking ahead,
the current implementation assumes a standardised 28-day cycle window for the
Fourier basis; future work should develop explicit methods for handling variable
cycle lengths.

\section{Conclusions}

\texttt{mcanalysis} provides a standardised, reproducible, open-source approach
to menstrual cycle analysis for users at all levels of statistical expertise,
encompassing the full pipeline from data preprocessing through statistical
modelling to publication-ready visualisation. Validation in a large dataset of
individuals with a range of chronic health conditions demonstrates evidence of
cycle effects across nine outcomes spanning physiological, sleep, symptom, mood,
and activity domains. The detectable cycle-related variation across these domains
suggests that cycle-aware adjustment of health recommendations and personalised
interventions may benefit users across a broad range of digital health
applications. The cyclic structure of the fitted models makes outputs directly
integrable into downstream predictive models and personalised intervention
frameworks such as dynamic treatment regimes, supporting future developments in
women's health research and personalised medicine. The package is freely
available in both R and Python at
\url{https://github.com/kyradelray/mcanalysis}, with a no-code web interface at
\url{https://kyradelray.shinyapps.io/mcanalysis/}.

\section*{Data Availability Statement}

The \texttt{mcanalysis} package, including example data and analysis scripts, is
freely available at \url{https://github.com/kyradelray/mcanalysis}. The Juli
dataset contains sensitive health information and is not publicly available.
Requests for data access should be directed to Juli Health.

\section*{Author Contributions}

K.D.\ conceived the study, developed the \texttt{mcanalysis} package (R and
Python implementations), performed all analyses, created all visualisations, and
wrote the manuscript. B.G.\ contributed to data interpretation and reviewed the
manuscript. G.L., J.H., and R.E.\ supervised the project and reviewed the
manuscript.

\section*{Competing Interests}

The authors declare no competing interests.

\bibliographystyle{unsrtnat}
\bibliography{references}

\end{document}